\documentclass[conference]{IEEEtran}
\usepackage[dvips]{graphicx}
\usepackage[cmex10]{amsmath}
\usepackage{multirow}
\usepackage{epsfig}
\usepackage{mathrsfs}
\usepackage{amssymb}
\usepackage{comment}
\usepackage{bm}
\usepackage{url}
\usepackage{array}
\usepackage{amsthm}
\usepackage{blkarray}
\usepackage{fancyhdr}
\usepackage{enumerate}
\usepackage[lined,boxed,commentsnumbered,ruled,linesnumbered]{algorithm2e}

\renewcommand{\thispagestyle}[1]{} 

\newtheorem{theorem}{Theorem}
\newtheorem{lemma}[theorem]{Lemma}

\newtheorem{corollary}[theorem]{Corollary}

\newcommand{\OmegaCAP}{\Omega^{\mbox{\tiny CAP}}}
\newcommand{\OmegaCSP}{\Omega^{\mbox{\tiny CSP}}}
\newcommand{\OmegaUP}{\Omega^{\mbox{\tiny UP}}}
\newcommand{\deltaMin}{\delta_{\mbox{\tiny min}}}
\newcommand{\MSC}{\mbox{MSC}}

\ifCLASSINFOpdf

\else

\fi

\usepackage[tight,footnotesize]{subfigure}

\begin{document}
\pagestyle{fancy}
\IEEEoverridecommandlockouts

\lhead{\textit{Technical Report, Dept. of EEE, Imperial College, London, UK, July, 2014.}}
\rhead{} 
%
\title{Fundamental Theories in Node Failure Localization}
\author{\IEEEauthorblockN{Liang Ma\IEEEauthorrefmark{2}, Ting He\IEEEauthorrefmark{3}, Ananthram Swami\IEEEauthorrefmark{4}, Don Towsley\IEEEauthorrefmark{1}, Kin K. Leung\IEEEauthorrefmark{2}, and Jessica Lowe\IEEEauthorrefmark{5}}
\IEEEauthorblockA{\IEEEauthorrefmark{2}Imperial College, London, UK. Email: \{l.ma10, kin.leung\}@imperial.ac.uk\\
\IEEEauthorrefmark{3}IBM T. J. Watson Research Center, Yorktown, NY, USA. Email: the@us.ibm.com\\
\IEEEauthorrefmark{4}Army Research Laboratory, Adelphi, MD, USA. Email: ananthram.swami.civ@mail.mil\\
\IEEEauthorrefmark{1}University of Massachusetts, Amherst, MA, USA. Email: towsley@cs.umass.eduzz\\
\IEEEauthorrefmark{5}DSTL, Salisbury, UK, jjlowe@dstl.gov.uk
}
}

\maketitle

\IEEEpeerreviewmaketitle

\section{Introduction}
Selected theorem proofs in \cite{Ma14AFMlong} are presented in detail in this report. We first list the theorems in Section II and then give the corresponding proofs in Section III. See the original paper \cite{Ma14AFMlong} for terms and definitions.

\section{Theorems}

\begin{lemma}[Abstract sufficient condition]\label{lem:abstract sufficient condition}
Any set of up to $k$ failed nodes is identifiable if for any non-monitor $v$ and failure set ${F}$ with $|{F}|\leq k$ ($v\not\in{F}$), there is a measurement path going through $v$ but no node in ${F}$.
\end{lemma}

\begin{lemma}[Abstract necessary condition]\label{lem:abstract necessary condition}
Any set of up to $k$ failed nodes is identifiable only if for any set $V'$ of non-monitors with $|V'|<k$, any set of up to $k-|V'|$ node failures is identifiable in $\mathcal{G}-V'$.
\end{lemma}

\begin{theorem}[$k$-identifiability under CAP]\label{thm:k-identifiability, CAP}
Network $\mathcal{G}$ is $k$-identifiable under CAP:
\begin{enumerate}
\item[a)] if for any set $V'$ of up to $k$ non-monitors, each connected component in $\mathcal{G}-V'$ contains a monitor;
\item[b)] only if for any set $V'$ of up to $k-1$ non-monitors, each connected component in $\mathcal{G}-V'$ contains a monitor.
\end{enumerate}
\end{theorem}

\begin{lemma}\label{lem:connectivity of G*}
Each connected component in $\mathcal{G}-V'$ contains a monitor for any set $V'$ of up to $s$ ($s\leq \sigma-1$) non-monitors if and only if $\mathcal{G}^*$ is $(s+1)$-vertex-connected.
\end{lemma}

\begin{corollary}\label{coro:sigma-identifiability, CAP}
Network $\mathcal{G}$ is $\sigma$-identifiable under CAP if and only if each non-monitor is the neighbor of a monitor.
\end{corollary}

\begin{theorem}[$k$-identifiability under CSP]\label{thm:k-identifiability, CSP}
Network $\mathcal{G}$ is $k$-identifiable under CSP:
\begin{enumerate}
\item[a)] if for any node set $V'$, $|V'|\leq k+1$, containing at most one monitor, each connected component in $\mathcal{G}-V'$ contains a monitor;
\item[b)] only if for any node set $V'$, $|V'|\leq k$, containing at most one monitor, each connected component in $\mathcal{G}-V'$ contains a monitor.
\end{enumerate}
\end{theorem}

\begin{lemma}\label{lem:connectivity of G_i}
The following two conditions are equivalent:
\begin{enumerate}
  \item[(1)] Each connected component in $\mathcal{G}-V'$ contains a monitor for any set $V'$ consisting of monitor $m$ ($m\in M$) and up to $s$ ($s\leq \sigma-1$) non-monitors;
  \item[(2)] $\mathcal{G}_m$ is $(s+1)$-vertex-connected.
\end{enumerate}
\end{lemma}

\begin{corollary}\label{coro:sigma-identifiability, CSP}
Network $\mathcal{G}$ is $\sigma$-identifiable under CSP if and only if each non-monitor has at least two monitors as neighbors.
\end{corollary}

\begin{corollary}\label{coro:(sigma-1)-identifiability, CSP}
Network $\mathcal{G}$ is $(\sigma-1)$-identifiable under CSP if and only if all but one non-monitor, denoted by $v$, have at least two monitors as neighbors, and $v$ either has (i) two or more monitors as neighbors, or (ii) one monitor and all the other non-monitors (i.e., $N\setminus \{v\}$) as neighbors.
\end{corollary}

\begin{theorem}[$k$-identifiability under UP]\label{thm:k-identifiability, UP}
Network $\mathcal{G}$ is $k$-identifiable under UP with measurement paths $P$: \begin{enumerate}
\item[a)] if $\MSC(v)>k$ for any non-monitor $v$;
\item[b)] only if $\MSC(v)>k-1$ for any non-monitor $v$.
\end{enumerate}
\end{theorem}

\begin{corollary}\label{coro:k-identifiability, CAP}
Network $\mathcal{G}$ is $k$-identifiable under CAP:
\begin{enumerate}
\item[a)] if $\mathcal{G}^*$ is $(k+1)$-vertex-connected ($k\leq \sigma-1$);
\item[b)] only if $\mathcal{G}^*$ is $k$-vertex-connected ($k\leq \sigma$).
\end{enumerate}
\end{corollary}

\begin{theorem}[Maximum Identifiability under CAP]\label{thm:Omega_CAP}
If $\delta(\mathcal{G}^*)\leq \sigma-1$, the maximum identifiability of $\mathcal{G}$ under CAP, $\OmegaCAP(\mathcal{G})$, is bounded by $\delta(\mathcal{G}^*) - 1 \leq \OmegaCAP(\mathcal{G}) \leq \delta(\mathcal{G}^*)$.
\end{theorem}

\begin{corollary}\label{coro:k-identifiability, CSP}
Network $\mathcal{G}$ is $k$-identifiable under CSP:
\begin{enumerate}
\item[a)] if $\mathcal{G}^*$ is $(k+2)$-vertex-connected, and $\mathcal{G}_m$ is $(k+1)$-vertex-connected for each monitor $m\in M$ ($k\leq \sigma-2$);
\item[b)] only if $\mathcal{G}^*$ is $(k+1)$-vertex-connected, and $\mathcal{G}_m$ is $k$-vertex-connected for each monitor $m\in M$ ($k\leq \sigma-1$).
\end{enumerate}
\end{corollary}

\begin{theorem}[Maximum Identifiability under CSP]\label{thm:Omega_CSP}
If $\min(\deltaMin,\: \delta(\mathcal{G}^*)-1) \leq \sigma-2$, the maximum identifiability of $\mathcal{G}$ under CSP, $\OmegaCSP(\mathcal{G})$, is bounded by $\min(\deltaMin-1,\: \delta(\mathcal{G}^*)-2) \leq \OmegaCSP(\mathcal{G}) \leq \min(\deltaMin,\: \delta(\mathcal{G}^*)-1)$.
\end{theorem}

\begin{theorem}[Maximum Identifiability under UP]\label{thm:Omega_UP}
The maximum identifiability of $\mathcal{G}$ under UP, $\OmegaUP(\mathcal{G})$, with measurement paths $P$ is bounded by $\Delta - 1 \leq \OmegaUP(\mathcal{G}) \leq \Delta$.
\end{theorem}

\section{Proofs}

\subsection{Proof of Lemma~\ref{lem:abstract sufficient condition}}

Consider two distinct failure sets ${F}$ and ${F}'$, each containing no more than $k$ nodes. There exists a node $v$ in only one of these sets; suppose $v\in {F}'\setminus F$. By the condition in the lemma, $\exists$ a path $p$ traversing $v$ but not ${F}$, thus distinguishing ${F}$ from ${F}'$.
\hfill$\blacksquare$

\subsection{Proof of Lemma~\ref{lem:abstract necessary condition}}

Suppose that $\exists$ two non-empty sets $V'$ and $V''$ of non-monitors, with $V'\cap V''=\emptyset$ and $|V'|+|V''|=k$, such that $V''$ is not identifiable in $\mathcal{G}-V'$. Then the union ${F}=V'\cup V''$ must be unidentifiable in $\mathcal{G}$, as even if we have identified failures in $V'$, we still cannot identify the rest of the failures.
\hfill$\blacksquare$

\subsection{Proof of Theorem~\ref{thm:k-identifiability, CAP}}

Suppose condition (a) holds, and consider a candidate failure set $V'$ and a non-monitor $v$ ($v\not\in V'$). Since the connected component in $\mathcal{G}-V'$ that contains $v$ has a monitor, there must exist a path connecting $v$ to a monitor that does not traverse any node in $V'$. Following this path from the monitor to $v$ and then back to the monitor then gives a path measurable under CAP that satisfies Lemma~\ref{lem:abstract sufficient condition}. Thus, condition (a) is sufficient.

Suppose condition (b) does not hold, i.e., there exists a non-monitor $v$ that is disconnected from all monitors in $\mathcal{G}-V'$ for a set $V'$ of up to $k-1$ non-monitors ($v\not\in V'$). Then if nodes in $V'$ fail, no remaining measurement path can probe $v$, and thus it is impossible to determine whether $v$ has failed or not. This violates the condition in Lemma~\ref{lem:abstract necessary condition}, and thus condition (b) is necessary.
\hfill$\blacksquare$

\subsection{Proof of Theorem~\ref{thm:k-identifiability, CAP}}

Suppose condition (a) holds, and consider a candidate failure set $V'$ and a non-monitor $v$ ($v\not\in V'$). Since the connected component in $\mathcal{G}-V'$ that contains $v$ has a monitor, there must exist a path connecting $v$ to a monitor that does not traverse any node in $V'$. Following this path from the monitor to $v$ and then back to the monitor then gives a path measurable under CAP that satisfies Lemma~\ref{lem:abstract sufficient condition}. Thus, condition (a) is sufficient.

Suppose condition (b) does not hold, i.e., there exists a non-monitor $v$ that is disconnected from all monitors in $\mathcal{G}-V'$ for a set $V'$ of up to $k-1$ non-monitors ($v\not\in V'$). Then if nodes in $V'$ fail, no remaining measurement path can probe $v$, and thus it is impossible to determine whether $v$ has failed or not. This violates the condition in Lemma~\ref{lem:abstract necessary condition}, and thus condition (b) is necessary.
\hfill$\blacksquare$

\subsection{Proof of Lemma~\ref{lem:connectivity of G*}}

We first show the equivalence between the first condition and the connectivity of $\mathcal{G}^*-V'$. If the first condition holds, then each connected component in $\mathcal{G}-M-V'$ contains a neighbor of a monitor. Since these neighbors are connected with each other and also with $m'$ in $\mathcal{G}^*-V'$, $\mathcal{G}^*-V'$ is connected. If the first condition is violated, i.e., there exists a connected component in $\mathcal{G}-M-V'$ without any neighbor of any monitor, then this component must be disconnected from $m'$, and hence $\mathcal{G}^*-V'$ must be disconnected.

We then show that requiring $\mathcal{G}^*-V'$ to be connected for any $V'$ of up to $s$ non-monitors is equivalent to requiring it to be connected for any $V'$ of up to $s$ nodes in $\mathcal{G}^*$, including $m'$, i.e., requiring $\mathcal{G}^*$ to be $(s+1)$-vertex-connected. It suffices to show that $\mathcal{G}^*-V'$ being connected for any $V'$ of up to $s$ non-monitors implies the connectivity of $\mathcal{G}^*-\{m'\}-V''$ for any $V''$ of up to $s-1$ non-monitors. Fixing a $V''$ of up to $s-1$ non-monitors, we assert that each connected component of $\mathcal{G}^*-\{m'\}-V''$ must contain a neighbor of a monitor, as otherwise $\mathcal{G}^*-V''$ will be disconnected. Since all these neighbors are connected via virtual links, $\mathcal{G}^*-\{m'\}-V''$ must be connected.
\hfill$\blacksquare$

\subsection{Proof of Corollary~\ref{coro:sigma-identifiability, CAP}}

If each non-monitor has a monitor as a neighbor, then their states can be determined independently through 1-hop probing, and hence any failure set is identifiable. On the other hand, if there exists a non-monitor $v$ that is only reachable by monitors via other non-monitors, then the state of $v$ cannot be determined in the case that all the other non-monitors fail, and hence $\mathcal{G}$ is not $\sigma$-identifiable.
\hfill$\blacksquare$

\subsection{Proof of Theorem~\ref{thm:k-identifiability, CSP}}

Suppose condition (a) holds, and consider a candidate failure set ${F}$, $|{F}|\leq k$ and a non-monitor $v\not\in {F}$. We argue that $v$ must have two simple \emph{vertex disjoint} paths to monitors in $\mathcal{G}-{F}$, and thus concatenating these paths provides a monitor-monitor simple path that traverses $v$ but not ${F}$, satisfying the abstract sufficient condition in Lemma~\ref{lem:abstract sufficient condition}. Indeed, if such paths do not exist, i.e., $\exists$ a (monitor or non-monitor) node $w$ ($w\neq v$) that resides on all paths from $v$ to monitors in $\mathcal{G}-{F}$, then $v$ will be disconnected from all monitors in $\mathcal{G}-{F}-\{w\}$, i.e., the connected component containing $v$ in $\mathcal{G}-V'$, where $V'={F}\cup \{w\}$, has no monitor, contradicting condition (a).

Suppose condition (b) does not hold, i.e., there exists a non-monitor $v$, a (monitor or non-monitor) node $w$, and a set of up to $k-1$ non-monitors ${F}$ ($v\neq w$ and $v,w\not\in{F}$) such that the connected component containing $v$ in $\mathcal{G}-V'$, $V'={F}\cup \{w\}$, contains no monitor. Then any path from $v$ to monitors in $\mathcal{G}-{F}$  must traverse $w$, which means no monitor-monitor simple path in $\mathcal{G}-{F}$ will traverse $v$ (as any monitor-monitor path traversing $v$ must form a cycle at $w$). This violates the necessary condition in Lemma~\ref{lem:abstract necessary condition} because if node $v$ fails, the failure cannot be identified in $\mathcal{G}-{F}$.\looseness=-1
\hfill$\blacksquare$

\subsection{Proof of Lemma~\ref{lem:connectivity of G_i}}

The proof is similar to that of Lemma~\ref{lem:connectivity of G*}. If the first condition holds, then each connected component in $\mathcal{G}-M-F$ for $F:= V'\setminus \{m\}$ contains a node in $\mathcal{N}(M\setminus \{m\})$, and thus $\mathcal{G}_m - F$ is connected. If the first condition is violated, then there is a connected component in $\mathcal{G}-M-F$ that does not contain any node in $\mathcal{N}(M\setminus \{m\})$. This component must be disconnected from $m'$ in $\mathcal{G}_m - F$, and thus $\mathcal{G}_m - F$ must be disconnected. Hence, the first condition is equivalent to $\mathcal{G}_m - F$ being connected for any set $F$ of up to $s$ non-monitors.
Moreover, $\mathcal{G}_m - F$ being connected for any set $F$ of up to $s$ non-monitors implies that $\mathcal{G}_m - \{m'\} - F'$ ($m'$ is the virtual monitor in $\mathcal{G}_m$) is connected for any $F'$ of up to $s-1$ non-monitors, because otherwise $\mathcal{G}_m - F'$ will be disconnected. Therefore, the first condition is equivalent to $\mathcal{G}_m - F$ being connected for any set $F$ of up to $s$ nodes in $\mathcal{G}_m$, i.e., the first and second conditions in Lemma~\ref{lem:connectivity of G_i} are equivalent.
\hfill$\blacksquare$

\subsection{Proof of Corollary~\ref{coro:sigma-identifiability, CSP}}

If each non-monitor has at least two monitors as neighbors, then their states can be determined independently by cycle-free 2-hop probing between monitors, and thus the network is $\sigma$-identifiable. On the other hand, suppose $\exists$ a non-monitor $v$ with zero or only one monitor neighbor. Then $\nexists$ simple paths going through $v$ without traversing another non-monitor, and hence the state of $v$ cannot be determined if all the other non-monitors fail.
\hfill$\blacksquare$

\subsection{Proof of Corollary~\ref{coro:(sigma-1)-identifiability, CSP}}

\emph{a) Necessity:} Suppose that $\mathcal{G}$ is $(\sigma-1)$-identifiable under CSP. If it is also $\sigma$-identifiable, then each non-monitor must have at least two monitors as neighbors according to Corollary~\ref{coro:sigma-identifiability, CSP}. Otherwise, we have $\Omega(\mathcal{G})=\sigma-1$. In this case, $\exists$ at least one non-monitor, denoted by $v$, with at most one monitor neighbor. Let $\mathcal{N}(v)$ denote all neighbors of $v$ including monitors. Suppose that $v$ has $\lambda$ neighbors (i.e., $|\mathcal{N}(v)|=\lambda$). Then there are two cases: (i) $\mathcal{N}(v)$ contains a monitor, denoted by $\widetilde{m}$; (ii) all nodes in $\mathcal{N}(v)$ are non-monitors. In case (i), the sets $F_1=\mathcal{N}(v)\setminus \{\widetilde{m}\}$ and $F_2=F_1\cup \{v\}$ are not distinguishable because $\nexists$ monitor-to-monitor simple paths traversing $v$ but not nodes in $F_1$. In case (ii), the sets $F_1=\mathcal{N}(v)\setminus \{w\}$ (where $w$ is an arbitrary node in $\mathcal{N}(v)$) and $F_2=F_1\cup \{v\}$ are not distinguishable as all monitor-to-monitor simple paths traversing $v$ must go through at least one node in $F_1$. Based on (i--ii), we conclude that $\Omega(\mathcal{G})\leq \lambda-1$, where $\lambda$ is the degree of any non-monitor with at most one monitor neighbor. For $\Omega(\mathcal{G})=\sigma-1$, we must have $\lambda\geq \sigma$, which can only be satisfied if all such non-monitors have one monitor and all the other non-monitors as neighbors. Moreover, if there are two such non-monitors $v$ and $u$, then the sets $F\cup \{v\}$ and $F\cup \{u\}$, where $F=N\setminus \{v,\: u\}$, are not distinguishable as all monitor-to-monitor simple paths traversing $v$ must go through $F$ or $u$ and vice versa. Therefore, such non-monitor must be unique.

\emph{b) Sufficiency:} If each non-monitor has at least two monitors as neighbors, then $\mathcal{G}$ is $\sigma$-identifiable (hence also $(\sigma-1)$-identifiable) according to Corollary~\ref{coro:sigma-identifiability, CSP}. If all but one non-monitor $v$ have at least two monitors as neighbors, and $v$ has one monitor $\widetilde{m}$ and all the other non-monitors (i.e., $N\setminus \{v\}$) as neighbors, then for any two failure sets $F_1$ and $F_2$ with $|F_i|\leq \sigma-1$ ($i=1,\: 2$), there are two cases: (i) $F_1$ and $F_2$ differ on a non-monitor other than $v$; (ii) $F_1$ and $F_2$ only differ on $v$. In case (i), since the states of all non-monitors other than $v$ can be independently determined, $F_1$ and $F_2$ are distinguishable. In case (ii), suppose that $F_1 = F\cup \{v\}$ and $F_2 = F$ for $F\subseteq N\setminus \{v\}$. Since $|F_1|\leq \sigma-1$, $|F|\leq \sigma-2$ and $\exists$ a non-monitor $w\in (N\setminus \{v\})\setminus F$. We know that $v$ is a neighbor of $w$ (as $v$ is a neighbor of all the other non-monitors) and $w$ is a neighbor of a monitor $m$ other than $\widetilde{m}$ (as it has at least two monitor neighbors). Thus, $\widetilde{m}vwm$ is a monitor-to-monitor simple path traversing $v$ but not $F$, whose measurement can distinguish $F_1$ and $F_2$. Therefore, $\mathcal{G}$ is $(\sigma-1)$-identifiable under CSP.
\hfill$\blacksquare$

\subsection{Proof of Theorem~\ref{thm:k-identifiability, UP}}

Suppose condition (a) holds. Then for any candidate failure set $F$ with $|F|\leq k$ and any other non-monitor $v$ ($v\not \in F$), there must be a path in $P_v$ that is not in $\bigcup_{w\in F}P_w$, i.e., traversing $v$ but not $F$, which satisfies the abstract sufficient condition in Lemma~\ref{lem:abstract sufficient condition}.

Suppose condition (b) does not hold, i.e., there exists a non-monitor $v$ and a set of non-monitors $V'$ with $|V'|\leq k-1$ and $v\not\in V'$, such that $P_v \subseteq \bigcup_{w\in V'}P_w$. Then given failures of all nodes in $V'$, the state of $v$ has no impact on observed path states and is thus unidentifiable, violating the abstract necessary condition in Lemma~\ref{lem:abstract necessary condition}.
\hfill$\blacksquare$

\subsection{Proof of Theorem~\ref{thm:Omega_CAP}}

Given $\delta(\mathcal{G}^*)$, we know that $\mathcal{G}^*$ is $\delta(\mathcal{G}^*)$-vertex-connected but not $(\delta(\mathcal{G}^*)+1)$-vertex-connected. By Corollary~\ref{coro:k-identifiability, CAP}, this means that $\mathcal{G}$ is $(\delta(\mathcal{G}^*)-1)$-identifiable but not $(\delta(\mathcal{G}^*)+1)$-identifiable, which yields the above bounds on the maximum identifiability. Note that applying Corollary~\ref{coro:k-identifiability, CAP} requires $\delta(\mathcal{G}^*)\leq \sigma-1$.
\hfill$\blacksquare$

\subsection{Proof of Theorem~\ref{thm:Omega_CSP}}

By definition of vertex-connectivity, $\mathcal{G}^*$ is $\delta(\mathcal{G}^*)$-vertex-connected, and $\mathcal{G}_m$ is $\deltaMin$-vertex-connected for each monitor $m \in M$. This satisfies the condition in Corollary~\ref{coro:k-identifiability, CSP}~(a) for $k = \min(\deltaMin-1,\: \delta(\mathcal{G}^*)-2)$, and thus $ \OmegaCSP(\mathcal{G}) \geq \min(\deltaMin-1,\: \delta(\mathcal{G}^*)-2)$. Meanwhile, since $\mathcal{G}^*$ is not $(\delta(\mathcal{G}^*)+1)$-vertex-connected, and $\mathcal{G}_m$ is not $(\deltaMin+1)$-vertex-connected for some $m \in M$, the condition in Corollary~\ref{coro:k-identifiability, CSP}~(b) is violated for $k = \min(\deltaMin+1,\: \delta(\mathcal{G}^*))$ (which requires $\min(\deltaMin+1,\: \delta(\mathcal{G}^*)) \leq \sigma-1$). Thus, $\OmegaCSP(\mathcal{G}) \leq \min(\deltaMin,\: \delta(\mathcal{G}^*)-1)$.
\hfill$\blacksquare$

\subsection{Proof of Theorem~\ref{thm:Omega_UP}}

Since $\MSC(v)>\Delta-1$ for all $v\in N$, $\mathcal{G}$ is $(\Delta-1)$-identifiable by Theorem~\ref{thm:k-identifiability, UP}~(a). Meanwhile, since there exists a node $v\in N$ with $\MSC(v) = \Delta$, $\mathcal{G}$ is not $(\Delta+1)$-identifiable by Theorem~\ref{thm:k-identifiability, UP}~(b). Together, they imply the bounds on $\OmegaUP(\mathcal{G})$.
\hfill$\blacksquare$

\bibliographystyle{IEEEtran}
\bibliography{mybibSimplifiedA}
\end{document}